\documentclass[a4paper]{article}
\usepackage[ansinew]{inputenc}
\usepackage{times}
\usepackage[T1]{fontenc}
\usepackage{graphicx}
\usepackage{geometry}
\usepackage{amssymb}
\usepackage{amsmath}

\usepackage{rotating}

\usepackage{xcolor}
\colorlet{shadecolor}{gray!25}
\usepackage{subfig}
\usepackage{float}
\usepackage{lscape}

\author{Ludwig A. Hothorn,\\ 
Im Grund 12, D-31867 Lauenau, Germany\\ \scriptsize(retired from Leibniz University Hannover)}

\title{Comparisons of multiple treatment groups with a negative control or placebo group: Dunnett test vs. closed test procedure}
\begin{document}

\maketitle
\begin{abstract}
Several treatments are usually compared with a control using the Dunnett test. As an alternative, three variants of the closed testing approach are considered, one with ANOVA-F-tests, one with MCT-GrandMean and one with global Dunnett-tests in the partition hypotheses.\\
\end{abstract}

\section{The problem}\label{sec1}
Comparisons of multiple treatment groups with a control (or placebo) group are frequently performed in biomedical experiments: in a plant molecular experiment new mutants were compared with the wild type \cite{Zhu2019}, in a toxicological assay selected clinical chemistry endpoints in three dose groups with disodium adenosine-triphosphate were compared  with a water control \cite{Jager2020} and in neurological clinical trial three candidate drugs amiloride, fluoxetine, and riluzole  were compared with a placebo group for the primary endpoint volumetric MRI percentage brain volume change \cite{Chataway2020}.\\
Typically, the Dunnett procedure is used \cite{Dunnett1955} in such randomized one-way designs. The main advantage is the availability of simultaneous confidence intervals, alternatively to the multiplicity-adjusted p-values. An alternative is the closed testing procedure (CTP) \cite{MARCUS1976}, because it is available for any test in the GLM (particularly its small $n_i$ approximations \cite{Schaarschmidt2009}) and can be easily generalized to further joint analysis, e.g.  considering multiple endpoints \cite{Bretz2009}. Three test versions are considered here for the global and partition hypotheses: the common ANOVA F-test, the global multiple contrast test for comparisons against the grand mean \cite{Pallmann2016} and the global Dunnett test. Since the last two are based on multiple contrast tests, subset contrasts can be formulated for all k-samples, allowing the entire degree of freedom to be used in all sub-tests.\\

%%%%%%%%%%%%%%%%%%%%%%%%%%%%%%%%%%%%%%%%%%%%%%%%%%%%%%%%%%%%%%%%%%%%%%%%%%%%%%%%%%%%%%%%%%%
\section{The Dunnett procedure}
In the following, a low-dimensional one-way layout $y_{ij}=\mu+ \text{factor}_i+\epsilon_{ij} (i=0,...,k)\  \text{with}\  \epsilon_{ij} \thicksim N(\mu, \sigma^2)$ is considered. The Dunnett test can be formulated as multiple contrast test (MCT) \cite{SHAFFER1977}, \cite{MUKERJEE1987}: $t_{MCT}=max(t_1,...,t_{q'})$ with $t_q=\sum_{i=0}^k c_i\bar{y}_i/S \sqrt{\sum_i^k c_i^2/n_i}$ where $c_i^q$ are the contrast coefficients (see below). The common-used adjusted p-values are given by the minimum empirical $\alpha$-level fulfilling the equality $\frac{\sum_{i=0}^k c_i\bar{y}_i}{S\sqrt{\sum_i^k c_i^2/n_i}} = t_{q,df,R,1-sided,1-min(\alpha)}$, where $t_{q,df,R,1-sided,1-\alpha}$ is the quantile of central q-variate t distribution, available in the package mvtnorm \cite{Mi2009}. Compatible to the adjusted p-values are (two) or one-sided simultaneous confidence limits which are not considered here because of their difficulties in the CTP \cite{Guilbaud2018, Westfall2013}.

%%%%%%%%%%%%%%%%%%%%%%%%%%%%%%%%%%%%%%%%%%%%%%%%%%%%%%%%%%
\section{The closed testing procedure for the comparison of treatment groups with a control}
An important issue of the CTP \cite{MARCUS1976} is the definition of the interesting elementary hypotheses, i.e. exactly those which are to be interpreted. Adjusted p-values are available for exactly these.  In the above described design, the many-to-one hypotheses $H_i: \mu_i-\mu_0$ are of interest only. Starting from this, one defines all subset intersection hypotheses up to the global hypothesis, involving these hypotheses. One rejects $H_i$ at level $\alpha$ if and only if $H_i$ itself is rejected and all hypotheses which include them (each at level  $\alpha$). Each hypothesis is tested with a level $\alpha$-test, with any appropriate test - this allows a high flexibility of the here described approach. Each of these tests (determined by the $\xi$ elementary hypotheses) is an intersection-union test (IUT), i.e. $T^{CTP}=min(T_1,...,T_{\xi})$, or more common $p^{CTP}=max(p_1,...,p_{\xi})$. In general CTP, the subset hypotheses can be complex and contradictory, but when considering hypotheses for comparisons with a control, they form a simple, so-called complete family of hypotheses \cite{Sonnemann2008} (same as with the 2-sample multiple endpoint problem). Using the monotonicity of the p-values and the dependencies of the sub-hypotheses, shortcuts are possible in general CTP \cite{Brannath2010}. By using additive or p-value combination tests, a variety of procedures can be constructed \cite{Henning2015} in the general case.\\
For the simple design with $k=2$, the family include the elementary (e.g. $H_0^{01}$), intersection (e.g. $H_0^{012}$) and global hypotheses (e.g. $H_0^{0123}$): \\
$H_0^{01}: \mu_0=\mu_1 \subset [H_0^{012}, H_0^{013}] \subset H_0^{0123}$\\
$H_0^{02}: \mu_0=\mu_2 \subset [H_0^{012}, H_0^{023}] \subset H_0^{0123}$\\
$H_0^{03}: \mu_0=\mu_3 \subset [H_0^{013}, H_0^{023}] \subset H_0^{0123}$\\

\subsection{Using ANOVA F-tests}
In the global, partition and elementary hypotheses the common omnibus F-test can be used: $F^{omnibus}=\frac{\sum_{i=0}^k(n_i(\bar{y}_i-\bar{\bar{y}})^2/k}{\sum_{i=0}^k\sum_j^{n_i}(n_i(y_{ij}-\bar{y}_i)^2/(N-k-1)}\thicksim F_{k,N-k-1}$. These global (partial) homogeneity null hypotheses $H^1: \mu_i=...= \mu_{k}$ are compared with $H_{F-test}^1: \mu_i\neq \mu_{i'}, \  \text{for at least one i, any i}, i\neq {i'}$ for $k(k-1)/2$ pairwise comparisons. This means that for the global hypothesis an F-test is used for all $(k+1)$ groups, for the first stage of the partition hypothesis an F-test is used for $(k+1-1)$ groups only, and so on, up to the elementary hypothesis an F-test with only 2 groups $(0,i)$. In the elementary hypotheses this loss of $df$ can be avoided by using a pairwise contrast test for $(k+1)$ groups (the so called multiple t-tests)- remember, one is free in the choice of the tests in CTP.

%%%%%%%%%%%%%%%%%%%%%%%%%%%%%%%%%%%%%%%%%%%%%%%%%%%%%%%%%%%%%%%%%%%%%%%%%
\subsection{Using global MCT for comparisons against grand mean}
The multiple contrast test against the grand mean (MCT-GM) represents an alternative to the F-test as global omnibus test \cite{Konietschke2013} although the alternative is different from those of the F-test: $H_{MCT-GM}^1: \mu_i\neq \bar{\bar{\mu}}$, i.e. only $k$ comparisons are considered. (Notice, this global test version ignores a key advantage of MCT-GM here: the availability of simultaneous confidence intervals.) The MCT-GM is a special case of a multiple contrast test, a maximum test: $t_{MCT}=max(t_1,...,t_q)$ where the $q$ contrast tests $	t_{Contrast}=\sum_{i=0}^k c_i\bar{y}_i/S \sqrt{\sum_i^k c_i^2/n_i}$ and the specific contrast coefficients (here again for k=3+1) in the design with a control (C) and three treatments $(T_{1,2,3})$:\\
	
				\begin{table}[ht]
				\centering\footnotesize
				\begin{tabular}{ c  c c r c c c c c }
         $c_i$ & C & $T_1$ & $T_2$ & $T_3$\\ \hline
        $c_a$ & -1 & 1/3   & 1/3 & 1/3   \\
        $c_b$ & 1/3 & -1 & 1/3 & 1/3  \\
				$c_c$ & 1/3 & 1/3 & -1 & 1/3  \\
				$c_d$ & 1/3 & 1/3 & 1/3&-1  \\
        \end{tabular}
				\caption{Contrast matrix for global grand mean comparisons}
        \end{table}
The here considered global MCT-GM is defined by $p_{MCT-GM}^{global}: max(p_{ii'})$. The advantage of this modified MCT-GM is the use of the full $df$for all hypotheses but subset contrasts for partition and elementary hypotheses. The contrast matrix for $H_0^{012}$ is for example:
	\begin{table}[ht]
	\centering\footnotesize
				\begin{tabular}{ c  c c r c c c c c}
         $c_i$ & C & $T_1$ & $T_2$ & $T_3$\\ \hline
        $c_a$ & -1 & 1/2   & 1/2 & 0  \\
        $c_b$ & 1/2 & -1 & 1/2 & 0 \\
				$c_c$ & 1/2 & 1/2 & -1 & 0 \\
        \end{tabular}
				\caption{Subset contrast matrix for grand mean comparisons for $H_0^{012}$}
        \end{table}

The  power advantage of MCT-GM with respect to the F-test can be observed for  small $n_i$ and several patterns of the alternative \cite{Pallmann2016}, where particularly the least favorable configuration (LFC)  $[0,0,...,0,\xi(k+1)/k]$ (effect size $\xi$) is of interest \cite{Konietschke2013}. A major advantage of the MCT's over the F-test is the easy availability of one-sided tests to avoid directional errors \cite{Westfall2013}.

%%%%%%%%%%%%%%%%%%%%%%%%%%%%%%%%%%%%%%%%%%%%%%%%%%%%%%%%%%%%%%%%%%%%%%%%%%%
\subsection{Using global Dunnett tests}

In the global and partition hypotheses global Dunnett test can be used: $p_{Dunnett}^{global}=max(p_1,...,p_q)$. Accordingly two-sample t-tests  are used in the elementary hypotheses. This power difference is a complex association between the $n_i$ (especially for unbalanced designs), the shape of the alternative, the effect size and the dimension $k$. Notice, in the elementary hypotheses again pairwise contrast tests for $(k+1)$ groups are used. Again, subset contrast matrices are used to keep the entire $df$ of the $k$-sample design.

\begin{table}[ht]
	\centering\footnotesize
				\begin{tabular}{ c  c c r c c c c  c}
         $c_i$ & C & $T_1$ & $T_2$ & $T_3$\\ \hline
				$c_a$ & -1 & 0 & 0  & 1   \\
        $c_b$ & -1 & 0 & 1  & 0   \\
        $c_c$ & -1 & 1 & 0  & 0  \\
				\end{tabular}
				\caption{Global Dunnett-type contrast matrix}
        \end{table}
Notice, the adjusted p-values for the Dunnett original procedure, the global Dunnett-test and the Grand Mean MCT were estimated by means of the package multcomp \cite{Hothorn2008}.

%%%%%%%%%%%%%%%%%%%%%%%%%%%%%%%%%%%%%%%%%%%%%%%%%%%%%%%%%%%%%%%%%%%%%%%%%%%%%%
\section{Simulation study}
In a simulation study for random experiments with a single primary endpoint $y_{ij}, k=3+1$, considering balanced and $n_0=sqrt(n_i)$ unbalanced one-way design and normal homoscedastic errors $\epsilon_{ij} \thicksim N(\mu, \sigma^2)$ selected alternatives, among them
$[\mu_0=\mu_1=\mu_2<\mu_3]; [\mu_0=\mu_1<\mu_2<\mu_3]; [\mu_0<\mu_1=\mu_2<\mu_3]; [\mu_0<\mu_1<\mu_2=\mu_3]$, were compared according to their FWER and any-pairs power  (5000/2000 runs). Common simulation studies on MCTs compare the any-pair power \cite{Hothorn2020} or average power \cite{Stevens2017} only. These concepts simplifies power comparisons considerably, but are not target-oriented, since which comparison is in the alternative is not considered. But one does not want to know if any mutants differ from the wild type, or any dose from the negative control, or any therapy vs. placebo. No, you want to evaluate exactly a particular mutant, dose or exactly a selected therapy  relative to control (see the motivating examples above). Therefore the concept of per-pairs power is shown in the Appendix, although it is difficult to interpret (and therefore k=3+1 was used). The four tests are abbreviated with D (Dunnett original), CTP-Du (subset Dunnett global), CTP-F(CTP ANOVA), and CTP-GM (subset CTP GrandMean) with $D_i$ the pairwise power, $D^a$ the any-pairs power. Instead complete power curves, only a relevant points in the alternative  with $\pi \approx 0.8$ is considered

\begin{table}[ht]
\centering\scriptsize
\begin{tabular}{l|l|rrrr}
  \hline
 $n_i$& $H_1$ &  Dunnett & CTP-Du & CTP-F &  CTP-GM \\ 
  \hline
 5,5,5,5 & $0,\delta,\delta, \delta$ &   0.87 &  0.91 &  0.86 & 0.88 \\ 
  & $0,0,0,\delta$ &   0.76 &  0.76 &  0.78 &  \textbf{0.83} \\ 
   & $0,\delta/3,2\delta/3, \delta$ &  0.76 & 0.78 &  0.70 & \textit{0.67} \\ 
   & $0,0,\delta, \delta$ &  0.76 &  0.87 & 0.86 &  0.81 \\ 
		& $0,\delta/5,\delta/3, \delta$ &  0.77 & 0.77 &  0.71 &  0.74 \\ 
		& $0,\delta/5,2\delta/3, \delta$ &  0.77 & 0.70 &  0.70 &  \textit{0.70} \\ 
	\hline
8,4,4,4 & $0,\delta,\delta, \delta$ &   0.79 &  0.95 &  0.86 & \textbf{0.96} \\ 
				& $0,0,0,\delta$ &   0.80 &  0.80 &  0.77 &  0.81 \\
				& $0,\delta/3,2\delta/3, \delta$ &  0.79 & 0.83 &  0.77 &  0.75 \\ 
				& $0,0,\delta, \delta$ &  0.76 &  0.87 & 0.86 &  0.81 \\ 
				& $0,\delta/5,\delta/3, \delta$ &  0.78 & 0.79 &  0.72 &  0.75 \\ 
				& $0,\delta/5,2\delta/3, \delta$ &  0.79 & 0.80 &  0.73 &  0.73 \\ \hline 
8,5,5,2	& $0,0,\delta, \delta$ &  0.53 &  0.88 & 0.88 & \textbf{ 0.84} \\ 		
	\hline	\hline 	
10,10,10,10 & $0,\delta,\delta, \delta$ &   0.91 &  0.95 &  0.91 & 0.92 \\ 
  & $0,0,0,\delta$ &   0.83 &  0.83 &  0.87 &  0.88 \\
	& $0,\delta/3,2\delta/3, \delta$ &  0.81 & 0.83 &  0.78 &  \textit{ 0.74} \\
	   & $0,0,\delta, \delta$ &  0.81 &  0.84 & 0.92 &  \textbf{0.88} \\ 
			& $0,\delta/5,\delta/3, \delta$ &  0.81 & 0.81 &  0.78 &  0.78 \\ 
			& $0,\delta/5,2\delta/3, \delta$ &  0.81 & 0.82 &  0.78 &  0.76 \\ 
	\hline
16,8,8,8 & $0,\delta,\delta, \delta$ &   0.94 &  0.97 &  0.97 & 0.97 \\
 				& $0,0,0,\delta$ &   0.84 &  0.84 &  0.83 &  0.86 \\
				& $0,\delta/3,2\delta/3, \delta$ &  0.84 & 0.87 &  0.83 &  0.80 \\ 
				& $0,0,\delta, \delta$ &  0.83 &  0.94 & 0.95 &  \textbf{0.90} \\ 
				& $0,\delta/5,\delta/3, \delta$ &  0.85 & 0.85 &  0.79 &  0.81 \\ 
				& $0,\delta/5,2\delta/3, \delta$ &  0.84 & 0.85 &  0.81 &  0.79 \\ 

 \hline  \hline
 20,20,20,20 & $0,\delta,\delta, \delta$ &   0.92 &  0.95 &  0.92 & 0.93 \\ 
  & $0,0,0,\delta$ &   0.83 &  0.84 &  0.86 &  0.88 \\
	   & $0,\delta/3,2\delta/3, \delta$ &  0.82 & 0.84 &  0.80 & \textit{0.75} \\
	   & $0,0,\delta, \delta$ &  0.83 &  0.93 & 0.93 & \textbf{0.90} \\ 
		& $0,\delta/5,\delta/3, \delta$ &  0.80 & 0.80 &  0.79 &  0.78 \\ 
		& $0,\delta/5,2\delta/3, \delta$ &  0.84 & 0.84 &  0.81 &  \textit{0.78} \\ 

   \hline
  38,14,14,14& $0,\delta,\delta, \delta$ &   0.95 &  0.98 &  0.98 & 0.98 \\ 
  & $0,0,0,\delta$ &   0.82 &  0.83 &  0.81 &  0.84 \\
	   & $0,\delta/3,2\delta/3, \delta$ &  0.82 & 0.87 &  0.84 &  0.83 \\
	   & $0,0,\delta, \delta$ &  0.83 &  0.93 & 0.93 & \textbf{ 0.90} \\ 
		& $0,\delta/5,\delta/3, \delta$ &  0.82 & 0.83 &  0.77 &  0.78 \\ 
		& $0,\delta/5,2\delta/3, \delta$ &  0.82 & 0.84 &  0.80 &  0.79 \\ 
		\hline

\end{tabular}
\caption{Any pairs power of the 4 tests \scriptsize(bold ... higher , italic... lower power in CTP-GM vs. Dunnett)}
\end{table}
\normalsize

Per definition all tests controls FWER empirically in both weak and strong sense (see the Appendix). Five tendencies reveal: i) no umpt exists, ii) the CTP-Du reveals  high power for $[\mu_0<\mu_1=...=\mu_{k-1}=\mu_k]$) alternative, iii) CTP-GM reveals high power for the $[\mu_0=\mu_1=...=\mu_{k-1}<\mu_k]$ alternative, i.e. the least favorable configuration of the Dunnett Test  \cite{Hayter1992}, iv) lowest power for CTP-GM is for alternatives where several groups contributes in proportion to the non-centrality, and v) for extreme unbalanced designs (e.g. $n_i=8,5,5,2$) and plateau shape the power loss of the Dunnett tests is clearly exceeded by the CTP tests.

%%%%%%%%%%%%%%%%%%%%%%%%%%%%%%%%%%%%%%%%%%%%%%%%%%%%%%%%%%%%%%
\section{Evaluation of a data example}
A randomized dose finding trial with 4 doses of a new compound (and placebo) to treat the irritable bowel syndrome measuring the primary efficacy endpoint baseline adjusted abdominal pain score was selected \cite{Biesheuvel2002}. (The raw data are available as data(IBScovars) in library(DoseFinding) \cite{DoseFinding}. 

\begin{figure}[ht]
	\centering
		\includegraphics[width=0.450\textwidth]{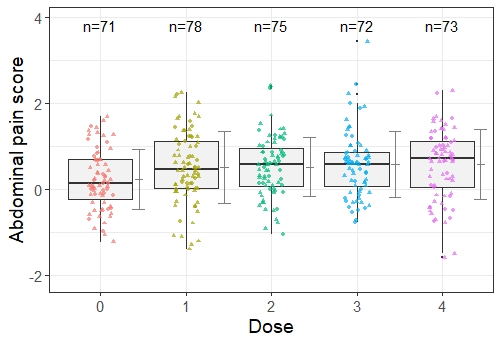}
	\caption{Boxplot for abdominal pain example}
	\label{fig:BoxplotEgbert}
\end{figure}

In the original work, the intention was a stratified analysis for males and females; here, gender is modeled as an additive factor without interaction. Although larger values of the endpoint represent a clinical benefit, two-sided tests are used for reasons of comparability.   
The two-sided multiplicity adjusted p-values for the 4 elementary hypotheses $\mu_i-\mu_0$ are given in Table 5.

\begin{table}[ht]
\centering\small
\begin{tabular}{r|l|lll}
 Comparison& Dunnett original  & CTP-F  & CTP-Du & CTP-GM  \\ \hline
 5-1& 0.0234 & 0.0346 & 0.0222 & 0.0121\\
 4-1& 0.0229  & 0.0346 & 0.0222 & 0.0117 \\
 3-1& 0.0654 & 0.0346 & 0.0358  & 0.0226\\ 
 2-1& 0.0779 & 0.0346 & 0.0358  & 0.0234\\
 \hline
\end{tabular}
\caption{Adjusted p-values for abdominal pain score}
\end{table}
The sensitivity advantages of all CTP's are obvious in this example. This even leads to qualitatively different statements: all doses are significantly different compared to placebo. The CTP-GM is particularly sensitive. The decision trees can be displayed particularly well with the CRAN package CPT \cite{CTP} in Figure 2.
\begin{figure}
	\centering
		\includegraphics[width=0.32\textwidth]{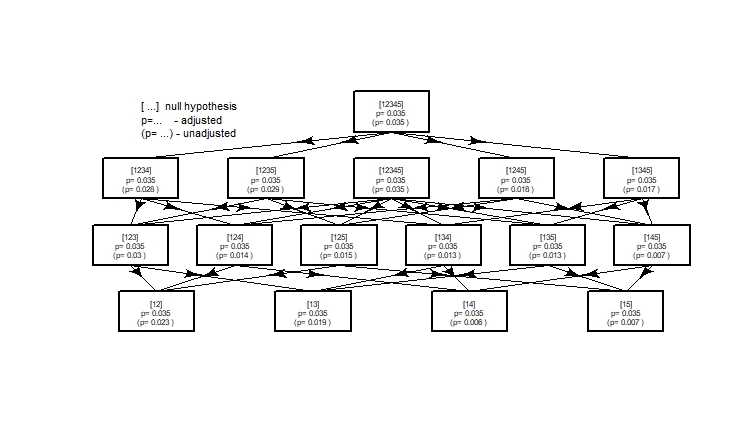}
		\includegraphics[width=0.32\textwidth]{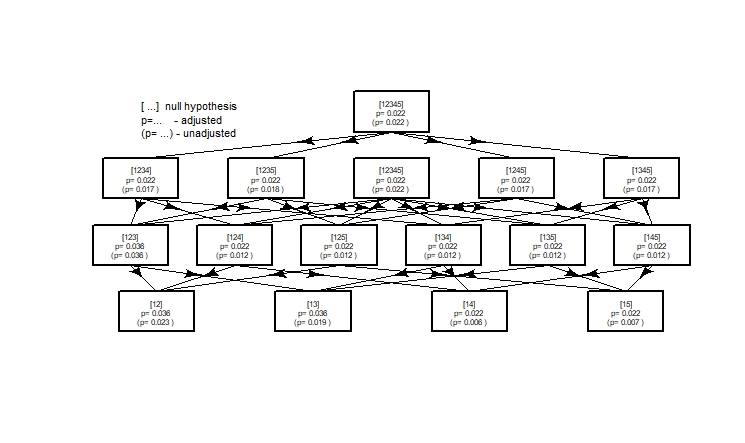}
		\includegraphics[width=0.32\textwidth]{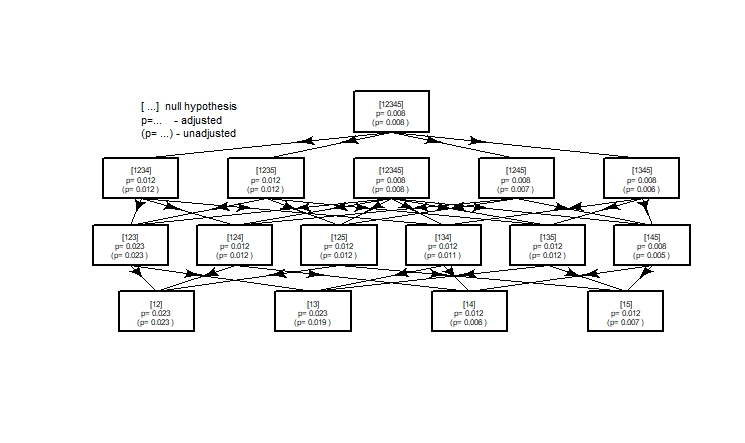}
	\caption{CTP's: (left) F-test, (middle) Dunnett global test, (right) GrandMeans MCT}
	\label{fig:CTPF}
\end{figure}

%%%%%%%%%%%%%%%%%%%%%%%%%%%%%%%%%%%%%%%%%%%%%%%%%%%%%%%%%%%%%%
\section{Summary}
Three modifications of the closed test procedure for comparisons of multiple treatment groups with a control as an alternative to the Dunnett procedure are proposed. As marginal tests the ANOVA F-Test, the global Dunnett test and the global test for Grand Mean multiple contrasts are considered. The disadvantage of unavailable confidence intervals is contrasted by the power advantage for selected shapes of the alternative, where of course none umpt exists. Related R-software is available on example style.\\
Extensions in the generalized linear (mixed) model, ratio-to-control inference, variance heterogeneity and non-parametric testing will be reported soon.

\footnotesize

\bibliographystyle{plain}
\footnotesize
%\bibliography{D:/PUB/Dunnett2020/Dunnett2020}

%\bibliographystyle{plain}
%\begin{thebibliography}{10}

%\bibitem{Bauer1996}
%P.~ Bauer and M.~ Kieser.
%\newblock A unifying approach for confidence intervals and testing of equivalence and difference.
%\newblock {\em Biometrika} 83: 934 -- 937, 1996.
%\end{thebibliography}
\normalsize
\section{Appendix}
%\newpage

\begin{landscape}
\begin{table}[ht]
\centering\tiny
\begin{tabular}{rrrrrrrrrrrrrrrrrrrrrrrrrrrr}
 & n1 & n2 & n3 & n4 & s1 & s2 & s3 & s4 & m2 & m3 & m4 & D1 & D2 & D3 & D &N1 & N2 & N3 & N & C1 & C2 & C3 & C & W1 & W2 & W3 & W \\
  \hline
1 & 5 & 5 & 5 & 5 & 1 & 1 & 1.4 & 1.4 & 10.0 & 10.0 & 10.0 & 0.020 & 0.019 & 0.022 & 0.039 & 0.022 & 0.020 & 0.024 & 0.051 & 0.014 & 0.014 & 0.014 & 0.033 & 0.015 & 0.015 & 0.014 & 0.034 \\
1 & 5 & 5 & 5 & 5 & 1 & 1 & 1.4 & 1.4 & 13.0 & 13.0 & 13.0 & 0.762 & 0.758 & 0.751 & 0.871 & 0.830 & 0.826 & 0.818 & 0.912 & 0.775 & 0.773 & 0.765 & 0.861 & 0.814 & 0.814 & 0.812 & 0.883 \\
1 & 5 & 5 & 5 & 5 & 1 & 1 & 1.4 & 1.4 & 10.0 & 10.0 & 13.0 & 0.018 & 0.013 & 0.752 & 0.757 & 0.028 & 0.021 & 0.754 & 0.761 & 0.025 & 0.020 & 0.772 & 0.779 & 0.028 & 0.022 & 0.825 & 0.833 \\
1 & 5 & 5 & 5 & 5 & 1 & 1 & 1.4 & 1.4 & 11.0 & 12.0 & 13.0 & 0.100 & 0.386 & 0.758 & 0.758 & 0.158 & 0.436 & 0.764 & 0.782 & 0.152 & 0.369 & 0.671 & 0.695 & 0.160 & 0.371 & 0.658 & 0.674 \\
1 & 5 & 5 & 5 & 5 & 1 & 1 & 1.4 & 1.4 & 10.0 & 13.0 & 13.0 & 0.018 & 0.762 & 0.755 & 0.758 & 0.040 & 0.789 & 0.778 & 0.874 & 0.050 & 0.780 & 0.771 & 0.863 & 0.050 & 0.750 & 0.744 & 0.809 \\
1 & 5 & 5 & 5 & 5 & 1 & 1 & 1.4 & 1.4 & 10.5 & 11.0 & 13.0 & 0.041 & 0.102 & 0.765 & 0.766 & 0.061 & 0.128 & 0.764 & 0.767 & 0.052 & 0.107 & 0.708 & 0.711 & 0.058 & 0.116 & 0.732 & 0.735 \\
1 & 5 & 5 & 5 & 5 & 1 & 1 & 1.4 & 1.4 & 10.5 & 11.5 & 13.0 & 0.029 & 0.203 & 0.765 & 0.767 & 0.057 & 0.244 & 0.770 & 0.775 & 0.053 & 0.213 & 0.694 & 0.702 & 0.060 & 0.216 & 0.689 & 0.695 \\
\hline
1 & 8 & 4 & 4 & 4 & 1 & 1 & 1.4 & 1.4 & 10.0 & 10.0 & 10.0 & 0.019 & 0.017 & 0.018 & 0.035 & 0.020 & 0.018 & 0.020 & 0.050 & 0.014 & 0.015 & 0.014 & 0.036 & 0.016 & 0.015 & 0.016 & 0.039 \\
1 & 8 & 4 & 4 & 4 & 1 & 1 & 1.4 & 1.4 & 13.0 & 13.0 & 13.0 & 0.789 & 0.785 & 0.785 & 0.910 & 0.854 & 0.860 & 0.860 & 0.953 & 0.850 & 0.854 & 0.858 & 0.943 & 0.866 & 0.874 & 0.872 & 0.957 \\
1 & 8 & 4 & 4 & 4 & 1 & 1 & 1.4 & 1.4 & 10.0 & 10.0 & 13.0 & 0.015 & 0.023 & 0.793 & 0.795 & 0.025 & 0.029 & 0.794 & 0.802 & 0.022 & 0.027 & 0.762 & 0.772 & 0.024 & 0.028 & 0.801 & 0.810 \\
1 & 8 & 4 & 4 & 4 & 1 & 1 & 1.4 & 1.4 & 11.0 & 12.0 & 13.0 & 0.102 & 0.399 & 0.784 & 0.790 & 0.174 & 0.464 & 0.796 & 0.827 & 0.172 & 0.415 & 0.728 & 0.768 & 0.183 & 0.426 & 0.725 & 0.750 \\
1 & 8 & 4 & 4 & 4 & 1 & 1 & 1.4 & 1.4 & 10.0 & 13.0 & 13.0 & 0.019 & 0.776 & 0.784 & 0.787 & 0.040 & 0.811 & 0.817 & 0.909 & 0.046 & 0.803 & 0.805 & 0.907 & 0.046 & 0.751 & 0.764 & 0.832 \\
1 & 8 & 4 & 4 & 4 & 1 & 1 & 1.4 & 1.4 & 10.5 & 11.0 & 13.0 & 0.028 & 0.090 & 0.776 & 0.778 & 0.052 & 0.122 & 0.780 & 0.785 & 0.050 & 0.105 & 0.711 & 0.719 & 0.058 & 0.108 & 0.739 & 0.745 \\
1 & 8 & 4 & 4 & 4 & 1 & 1 & 1.4 & 1.4 & 10.5 & 11.5 & 13.0 & 0.035 & 0.207 & 0.787 & 0.787 & 0.064 & 0.254 & 0.789 & 0.797 & 0.068 & 0.230 & 0.721 & 0.734 & 0.071 & 0.238 & 0.720 & 0.728 \\
1 & 8 & 5 & 5 & 2 & 1 & 1 & 1.4 & 1.4 & 10.0 & 10.0 & 10.0 & 0.018 & 0.016 & 0.018 & 0.033 & 0.018 & 0.017 & 0.019 & 0.047 & 0.015 & 0.013 & 0.014 & 0.036 & 0.016 & 0.015 & 0.017 & 0.042 \\
1 & 8 & 5 & 5 & 2 & 1 & 1 & 1.4 & 1.4 & 13.0 & 13.0 & 13.0 & 0.849 & 0.840 & 0.532 & 0.891 & 0.886 & 0.877 & 0.671 & 0.949 & 0.871 & 0.865 & 0.670 & 0.932 & 0.894 & 0.894 & 0.680 & 0.948 \\
1 & 8 & 5 & 5 & 2 & 1 & 1 & 1.4 & 1.4 & 10.0 & 10.0 & 13.0 & 0.020 & 0.017 & 0.525 & 0.533 & 0.028 & 0.027 & 0.526 & 0.537 & 0.030 & 0.027 & 0.469 & 0.480 & 0.031 & 0.027 & 0.533 & 0.544 \\
1 & 8 & 5 & 5 & 2 & 1 & 1 & 1.4 & 1.4 & 11.0 & 12.0 & 13.0 & 0.106 & 0.456 & 0.542 & 0.568 & 0.172 & 0.500 & 0.575 & 0.695 & 0.171 & 0.457 & 0.514 & 0.639 & 0.181 & 0.449 & 0.500 & 0.610 \\
1 & 8 & 5 & 5 & 2 & 1 & 1 & 1.4 & 1.4 & 10.0 & 13.0 & 13.0 & 0.018 & 0.832 & 0.527 & 0.534 & 0.040 & 0.849 & 0.581 & 0.875 & 0.050 & 0.847 & 0.536 & 0.880 & 0.052 & 0.818 & 0.551 & 0.842 \\
1 & 8 & 5 & 5 & 2 & 1 & 1 & 1.4 & 1.4 & 10.5 & 11.0 & 13.0 & 0.034 & 0.118 & 0.533 & 0.538 & 0.060 & 0.150 & 0.542 & 0.559 & 0.061 & 0.136 & 0.452 & 0.472 & 0.066 & 0.131 & 0.463 & 0.481 \\
1 & 8 & 5 & 5 & 2 & 1 & 1 & 1.4 & 1.4 & 10.5 & 11.5 & 13.0 & 0.046 & 0.245 & 0.517 & 0.525 & 0.074 & 0.284 & 0.533 & 0.593 & 0.072 & 0.263 & 0.457 & 0.519 & 0.078 & 0.252 & 0.440 & 0.493 \\
\hline
1 & 8 & 5 & 5 & 2 & 1 & 1 & 1.4 & 1.4 & 10.0 & 10.0 & 10.0 & 0.018 & 0.016 & 0.018 & 0.033 & 0.018 & 0.017 & 0.019 & 0.047 & 0.015 & 0.013 & 0.014 & 0.036 & 0.016 & 0.015 & 0.017 & 0.042 \\
1 & 8 & 5 & 5 & 2 & 1 & 1 & 1.4 & 1.4 & 13.0 & 13.0 & 13.0 & 0.849 & 0.840 & 0.532 & 0.891 & 0.886 & 0.877 & 0.671 & 0.949 & 0.871 & 0.865 & 0.670 & 0.932 & 0.894 & 0.894 & 0.680 & 0.948 \\
1 & 8 & 5 & 5 & 2 & 1 & 1 & 1.4 & 1.4 & 10.0 & 10.0 & 13.0 & 0.020 & 0.017 & 0.525 & 0.533 & 0.028 & 0.027 & 0.526 & 0.537 & 0.030 & 0.027 & 0.469 & 0.480 & 0.031 & 0.027 & 0.533 & 0.544 \\
1 & 8 & 5 & 5 & 2 & 1 & 1 & 1.4 & 1.4 & 11.0 & 12.0 & 13.0 & 0.106 & 0.456 & 0.542 & 0.568 & 0.172 & 0.500 & 0.575 & 0.695 & 0.171 & 0.457 & 0.514 & 0.639 & 0.181 & 0.449 & 0.500 & 0.610 \\
1 & 8 & 5 & 5 & 2 & 1 & 1 & 1.4 & 1.4 & 10.0 & 13.0 & 13.0 & 0.018 & 0.832 & 0.527 & 0.534 & 0.040 & 0.849 & 0.581 & 0.875 & 0.050 & 0.847 & 0.536 & 0.880 & 0.052 & 0.818 & 0.551 & 0.842 \\
1 & 8 & 5 & 5 & 2 & 1 & 1 & 1.4 & 1.4 & 10.5 & 11.0 & 13.0 & 0.034 & 0.118 & 0.533 & 0.538 & 0.060 & 0.150 & 0.542 & 0.559 & 0.061 & 0.136 & 0.452 & 0.472 & 0.066 & 0.131 & 0.463 & 0.481 \\
1 & 8 & 5 & 5 & 2 & 1 & 1 & 1.4 & 1.4 & 10.5 & 11.5 & 13.0 & 0.046 & 0.245 & 0.517 & 0.525 & 0.074 & 0.284 & 0.533 & 0.593 & 0.072 & 0.263 & 0.457 & 0.519 & 0.078 & 0.252 & 0.440 & 0.493 \\ 
   \hline\hline

1 & 10 & 10 & 10 & 10 & 2 & 2 & 2.0 & 2.0 & 10.0 & 10.0 & 10.0 & 0.016 & 0.022 & 0.018 & 0.033 & 0.019 & 0.024 & 0.021 & 0.050 & 0.013 & 0.014 & 0.014 & 0.033 & 0.014 & 0.016 & 0.015 & 0.034 \\
1 & 10 & 10 & 10 & 10 & 2 & 2 & 2.0 & 2.0 & 10.5 & 11.0 & 13.0 & 0.035 & 0.107 & 0.806 & 0.806 & 0.056 & 0.138 & 0.806 & 0.807 & 0.054 & 0.117 & 0.779 & 0.783 & 0.056 & 0.115 & 0.780 & 0.783 \\
1 & 10 & 10 & 10 & 10 & 2 & 2 & 2.0 & 2.0 & 10.5 & 11.5 & 13.0 & 0.038 & 0.227 & 0.809 & 0.812 & 0.070 & 0.276 & 0.813 & 0.818 & 0.067 & 0.243 & 0.769 & 0.777 & 0.072 & 0.245 & 0.752 & 0.758 \\
1 & 10 & 10 & 10 & 10 & 2 & 2 & 2.0 & 2.0 & 13.0 & 13.0 & 13.0 & 0.801 & 0.807 & 0.811 & 0.914 & 0.868 & 0.863 & 0.863 & 0.948 & 0.840 & 0.844 & 0.845 & 0.914 & 0.860 & 0.860 & 0.855 & 0.922 \\
1 & 10 & 10 & 10 & 10 & 2 & 2 & 2.0 & 2.0 & 10.0 & 13.0 & 13.0 & 0.018 & 0.809 & 0.804 & 0.811 & 0.037 & 0.842 & 0.835 & 0.926 & 0.045 & 0.845 & 0.847 & 0.920 & 0.045 & 0.825 & 0.818 & 0.881 \\
1 & 10 & 10 & 10 & 10 & 2 & 2 & 2.0 & 2.0 & 11.0 & 12.0 & 13.0 & 0.109 & 0.448 & 0.808 & 0.809 & 0.177 & 0.506 & 0.818 & 0.834 & 0.169 & 0.451 & 0.754 & 0.778 & 0.178 & 0.433 & 0.728 & 0.742 \\
1 & 10 & 10 & 10 & 10 & 2 & 2 & 2.0 & 2.0 & 10.0 & 10.0 & 13.0 & 0.027 & 0.020 & 0.824 & 0.829 & 0.034 & 0.034 & 0.827 & 0.833 & 0.032 & 0.032 & 0.865 & 0.873 & 0.035 & 0.031 & 0.875 & 0.881 \\ 
\hline
1 & 16 & 8 & 8 & 8 & 2 & 2 & 2.0 & 2.0 & 10.0 & 10.0 & 10.0 & 0.018 & 0.018 & 0.018 & 0.034 & 0.019 & 0.019 & 0.019 & 0.048 & 0.013 & 0.015 & 0.015 & 0.035 & 0.014 & 0.016 & 0.017 & 0.040 \\
1 & 16 & 8 & 8 & 8 & 2 & 2 & 2.0 & 2.0 & 10.5 & 11.0 & 13.0 & 0.042 & 0.107 & 0.845 & 0.846 & 0.071 & 0.140 & 0.846 & 0.850 & 0.068 & 0.120 & 0.785 & 0.790 & 0.069 & 0.128 & 0.805 & 0.808 \\
1 & 16 & 8 & 8 & 8 & 2 & 2 & 2.0 & 2.0 & 10.5 & 11.5 & 13.0 & 0.042 & 0.238 & 0.835 & 0.836 & 0.071 & 0.292 & 0.842 & 0.851 & 0.073 & 0.260 & 0.794 & 0.810 & 0.075 & 0.257 & 0.774 & 0.786 \\
1 & 16 & 8 & 8 & 8 & 2 & 2 & 2.0 & 2.0 & 13.0 & 13.0 & 13.0 & 0.825 & 0.831 & 0.843 & 0.942 & 0.890 & 0.892 & 0.902 & 0.968 & 0.890 & 0.896 & 0.911 & 0.966 & 0.901 & 0.902 & 0.912 & 0.972 \\
1 & 16 & 8 & 8 & 8 & 2 & 2 & 2.0 & 2.0 & 10.0 & 13.0 & 13.0 & 0.022 & 0.840 & 0.823 & 0.827 & 0.051 & 0.861 & 0.855 & 0.943 & 0.056 & 0.866 & 0.856 & 0.946 & 0.056 & 0.838 & 0.831 & 0.904 \\
1 & 16 & 8 & 8 & 8 & 2 & 2 & 2.0 & 2.0 & 11.0 & 12.0 & 13.0 & 0.103 & 0.412 & 0.835 & 0.838 & 0.171 & 0.475 & 0.844 & 0.867 & 0.171 & 0.450 & 0.795 & 0.827 & 0.178 & 0.445 & 0.782 & 0.802 \\
1 & 16 & 8 & 8 & 8 & 2 & 2 & 2.0 & 2.0 & 10.0 & 10.0 & 13.0 & 0.020 & 0.017 & 0.835 & 0.837 & 0.030 & 0.023 & 0.835 & 0.840 & 0.025 & 0.021 & 0.828 & 0.833 & 0.032 & 0.024 & 0.856 & 0.862 \\ 
\hline \hline
1 & 20 & 20 & 20 & 20 & 3 & 3 & 2.9 & 2.9 & 10.0 & 10.0 & 10.0 & 0.019 & 0.017 & 0.021 & 0.037 & 0.020 & 0.018 & 0.022 & 0.051 & 0.014 & 0.014 & 0.015 & 0.035 & 0.015 & 0.013 & 0.014 & 0.035 \\
1 & 20 & 20 & 20 & 20 & 3 & 3 & 2.9 & 2.9 & 10.0 & 10.0 & 13.0 & 0.018 & 0.019 & 0.826 & 0.832 & 0.026 & 0.024 & 0.827 & 0.836 & 0.025 & 0.024 & 0.858 & 0.867 & 0.028 & 0.025 & 0.867 & 0.875 \\
1 & 20 & 20 & 20 & 20 & 3 & 3 & 2.9 & 2.9 & 13.0 & 13.0 & 13.0 & 0.829 & 0.818 & 0.817 & 0.923 & 0.879 & 0.868 & 0.875 & 0.953 & 0.857 & 0.849 & 0.855 & 0.919 & 0.872 & 0.862 & 0.869 & 0.932 \\
1 & 20 & 20 & 20 & 20 & 3 & 3 & 2.9 & 2.9 & 11.0 & 12.0 & 13.0 & 0.100 & 0.428 & 0.814 & 0.817 & 0.169 & 0.488 & 0.822 & 0.844 & 0.166 & 0.442 & 0.773 & 0.796 & 0.172 & 0.420 & 0.733 & 0.754 \\
1 & 20 & 20 & 20 & 20 & 3 & 3 & 2.9 & 2.9 & 10.0 & 13.0 & 13.0 & 0.018 & 0.818 & 0.823 & 0.829 & 0.041 & 0.844 & 0.844 & 0.927 & 0.050 & 0.858 & 0.864 & 0.933 & 0.050 & 0.836 & 0.839 & 0.898 \\
1 & 20 & 20 & 20 & 20 & 3 & 3 & 2.9 & 2.9 & 10.5 & 11.0 & 13.0 & 0.035 & 0.093 & 0.803 & 0.804 & 0.050 & 0.128 & 0.804 & 0.805 & 0.052 & 0.112 & 0.783 & 0.786 & 0.052 & 0.105 & 0.774 & 0.778 \\
1 & 20 & 20 & 20 & 20 & 3 & 3 & 2.9 & 2.9 & 10.5 & 11.5 & 13.0 & 0.037 & 0.242 & 0.832 & 0.835 & 0.066 & 0.295 & 0.836 & 0.843 & 0.066 & 0.261 & 0.800 & 0.812 & 0.068 & 0.245 & 0.774 & 0.785 \\
\hline
1 & 38 & 14 & 14 & 14 & 3 & 3 & 2.9 & 2.9 & 10.0 & 10.0 & 10.0 & 0.018 & 0.015 & 0.017 & 0.034 & 0.019 & 0.015 & 0.018 & 0.047 & 0.014 & 0.012 & 0.012 & 0.033 & 0.016 & 0.013 & 0.012 & 0.036 \\
1 & 38 & 14 & 14 & 14 & 3 & 3 & 2.9 & 2.9 & 10.0 & 10.0 & 13.0 & 0.015 & 0.018 & 0.820 & 0.823 & 0.020 & 0.025 & 0.821 & 0.828 & 0.023 & 0.023 & 0.798 & 0.805 & 0.021 & 0.025 & 0.832 & 0.839 \\
1 & 38 & 14 & 14 & 14 & 3 & 3 & 2.9 & 2.9 & 13.0 & 13.0 & 13.0 & 0.820 & 0.817 & 0.827 & 0.948 & 0.890 & 0.882 & 0.892 & 0.978 & 0.898 & 0.888 & 0.899 & 0.979 & 0.897 & 0.893 & 0.902 & 0.981 \\
1 & 38 & 14 & 14 & 14 & 3 & 3 & 2.9 & 2.9 & 11.0 & 12.0 & 13.0 & 0.098 & 0.429 & 0.815 & 0.821 & 0.175 & 0.495 & 0.836 & 0.872 & 0.181 & 0.466 & 0.802 & 0.843 & 0.187 & 0.468 & 0.789 & 0.827 \\
1 & 38 & 14 & 14 & 14 & 3 & 3 & 2.9 & 2.9 & 10.0 & 13.0 & 13.0 & 0.017 & 0.819 & 0.815 & 0.818 & 0.042 & 0.848 & 0.844 & 0.946 & 0.050 & 0.849 & 0.841 & 0.949 & 0.050 & 0.832 & 0.821 & 0.912 \\
1 & 38 & 14 & 14 & 14 & 3 & 3 & 2.9 & 2.9 & 10.5 & 11.0 & 13.0 & 0.035 & 0.112 & 0.815 & 0.820 & 0.052 & 0.147 & 0.818 & 0.826 & 0.051 & 0.130 & 0.767 & 0.773 & 0.055 & 0.134 & 0.776 & 0.782 \\
1 & 38 & 14 & 14 & 14 & 3 & 3 & 2.9 & 2.9 & 10.5 & 11.5 & 13.0 & 0.034 & 0.209 & 0.818 & 0.820 & 0.056 & 0.259 & 0.825 & 0.843 & 0.062 & 0.239 & 0.773 & 0.797 & 0.066 & 0.242 & 0.775 & 0.793 \\
   \hline
\end{tabular}
\caption{FWER and per-pairs power. \footnotesize D (Dunnett original), N (Dunnett global), C(CTP ANOVA), W(CTP MCT-GrandMean)}
\end{table}
\end{landscape}

\end{document}